\documentstyle[aps,prb,multicol,epsf]{revtex}

\begin{document}
\draft

%\twocolumn[\hsize\textwidth\columnwidth\hsize
%\csname @twocolumnfalse\endcsname

\title{Anisotropic Strong Coupling Calculation of the Local
Electromagnetic Response of High-T$_c$ Superconductors}
\author{A. Bille and K. Scharnberg}
\address{Fachbereich Physik, Universit\"at Hamburg,
Jungiusstra\ss e 11, D-20355 Hamburg, Germany}

%\date{\today}
\maketitle

\begin{abstract}
The electromagnetic response of the CuO$_2$-planes is calculated within a strong
coupling theory using model tight binding bands and momentum dependent
pairing interactions representing spin fluctuation and phonon exchange.
The superconducting state resulting from these interactions has $d$-wave
symmetry. With phonon exchange included the order parameter amplitude
grows rapidly below T$_c$ at elevated frequencies  which leads to improved
agreement with the observed temperature dependence of the penetration
depth. Good agreement between theory and experiment can only be achieved
if it is assumed that the strength of
the quasiparticle interaction decreases with temperature in the superconducting
state. The amount of this reduction depends sensitively on the momentum
dependence of the interactions, the energy dispersion and the position of the
Fermi line.
\end{abstract}

%\section{INTRODUCTION}
%\label{sec:intro}
\begin{multicols}{2}
The fundamental information required to perform strong coupling calculations
for conventional superconductors is the Eliashberg function which incorporates
both the phonon spectrum and the electron-phonon interaction. These
properties can be determined in the
normal state and they are not expected to to be affected significantly
by the onset of superconductivity. Since much evidence for a highly anisotropic
pair state in the high-T$_c$ materials has accumulated in the recent past,
the momentum dependence of the Eliashberg function needs to be taken into
account. Upon the assumption that the underlying interaction is an exchange of
spin-fluctuations, such an Eliashberg function has been modelled quite
successfully to explain normal state properties. There is still some controversy
about the best possible model
but we feel that the function suggested by Monthoux and Pines,
\cite{Pinesa}
which contains the momentum dependence in a nonseparable form, is a reasonable
starting point.
%The big problem with this choice for an Eliashberg function is that,
Since this Eliashberg function is supposed to arise
from electron-electron interactions,
we expect it to change when superconductivity sets in.
One way this change could manifest itself is a lower frequency cut-off
approximately of the size and temperature dependence of the superconducting
order parameter amplitude. \cite{Schachinger97}
%In view of the $d$-wave state anticipated a change in the frequency dependence
%of the density of states from linear to cubic for frequencies below the order
%parameter amplitude, as used in the nested Fermi liquid model, \cite{Ruvalds}
%might be a better approximation.

In this paper we have not included such strongly temperature dependent
modifications of the Eliashberg function.
%\cite{Schachinger97,Ruvalds} in agreement with experiment.
%\cite{Bonna,Bonnb,Hensen97}
Instead, we focus on the momentum and frequency
dependence of the various self-energies,
in particular the scattering rate, which in conjunction with the anisotropy
of the Fermi velocity could have a marked effect on the conductivity.
For an energy dispersion of the form \cite{Schneider}
\begin{eqnarray}
\varepsilon ({\bf k}) &=& -t\left\lbrack 2\cos (k_x\pi ) + 2\cos (k_y\pi )
\right. \cr
%& &\left.\right.\cr
& & ~~~~~~~~~~\left. - 4B \cos (k_x\pi )\cos (k_y\pi ) + \mu\right\rbrack ,
\label{qpdisp}
\end{eqnarray}
%}\hfill

\noindent
with $k_{x,y}$ given in units of $\pi /a$,
we have plotted in Fig. 1 the Fermi lines (inset) and the Fermi velocities in
the first quadrant of the Brillouin zone.
For each choice of the parameter
$B$ the band filling parameter $\mu$ is chosen such that
the Fermi line meets the zone boundary at the same point $(1,0.1)$.
For a simple tight binding band $B=0$, quasiparticles with momenta in $(1,1)$
direction have indeed the highest Fermi velocities and thus contribute most
strongly to the conductivity.  For the modified tight
binding band with $B=0.45$  there is very little variation of the Fermi
velocity, which may be an artefact of this particular model band structure,
as the comparison with the dot-dot-dashed curves show.
\epsfxsize=3.20in
\begin{figure}[hb]
  \vspace{-1.5cm}
% \hspace{0.1cm}
  \label{fig1}
%  \begin{center}
    \centerline{%
    \begin{minipage}{8.5cm}
       \mbox{\epsffile{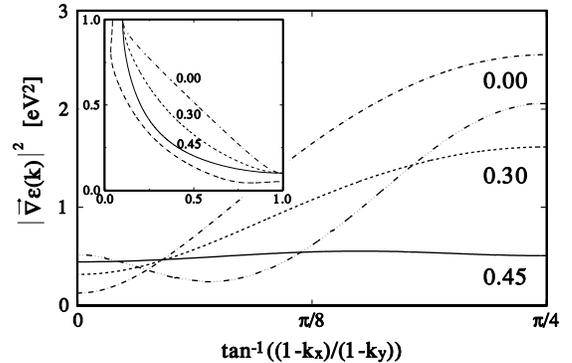}}\
       \parbox{6cm}{\caption{Fermi velocities $\vert\vec\nabla
\varepsilon ({\bf k})\vert^2 = v_F^2\,\left(\hbar\pi/a\right)^2$ as calculated
from Eq.\ ({\protect\ref{qpdisp}}) for a band width $8t = 1.44\,$eV.
%The parameter sets $(B,\mu )$ are $(0.00,0.098)$, $(0.30,-1.043)$,
%and $(0.45,-1.614)$, respectively.
The inset shows the corresponding Fermi lines.
The dot-dot-dashed lines are for the 8 parameter tight binding approximation to
the odd band of the CuO$_2$-double layers in YBCO,
{\protect\cite{OKA}} scaled to the same band width.
}}
     \end{minipage}
    }
%  \end{center}
\end{figure}
  \vspace{-0.2cm}

It has been noted before that a pairing interaction,
which is attractive at small momentum transfers,
can contribute to the formation of a $d$-wave state.
\cite{Dahm}
In addition to the spin fluctuation induced pairing interaction \cite{Pinesa}
  \vspace{-0.1cm}
\begin{equation}
\alpha^2\!F_{SF}({\bf q},\Omega ) = {g_{SF}^2\over \pi}{\chi_Q{\Omega
\over\omega_{SF}}\over
\Big[ 1\!+\!\pi^2\xi^2
\left({\bf q}\!-\!{\bf Q}\right)^2\big]^2 +
{\Omega^2\over \omega_{SF}^2}}
\label{elisf}
\end{equation}
  \vspace{-0.1cm}

\noindent
we have, therefore incorporated into our strong coupling calculations
a phonon contribution of the form

  \vspace{-0.1cm}
%\begin{eqnarray}
%\alpha^2F_{ph}{\bf q},\Omega ) &=&
%g_{ph}\,{1\over \pi} G({\bf q})\,\bigg(
%{\Gamma^3\Omega\over \lbrack\left(\Omega
%-\Omega_0\right)^2+\Gamma^2\rbrack^2}\cr
%&  & \cr
%&&~~~~~~~~~~~~~~+ {\Gamma^3\Omega\over \lbrack\left(\Omega
%+\Omega_0\right)^2+\Gamma^2\rbrack^2} \bigg)\,.
%\label{eliph}
%\end{eqnarray}
\begin{equation}
\alpha^2F_{ph}({\bf q},\Omega ) =
g_{ph}\,{1\over \pi} G({\bf q}) \sum_{\sigma=\pm}
{\Gamma^3\Omega\over \lbrack\left(\Omega
+\sigma \Omega_0\right)^2+\Gamma^2\rbrack^2}
\label{eliph}
\end{equation}
Numerical results are for $\Omega_0 = 41\,$meV	and $\Gamma = 8\,$meV.
A very similar frequency dependence for $\alpha^2F_{ph}$ has been suggested
by Golubov.\cite{Golubov}
Note that at low frequencies both functions vary linearly with frequency
but for a typical value $\omega_{SF} = 7.7\,$meV, \cite{Pinesa}
which is much smaller than the value $30\,$meV used in Ref.\
[\onlinecite{Schachinger97}],
$\alpha^2F_{SF}$ dominates.

We have antisymmetrized $\alpha^2F_{ph}({\bf q},\Omega )$ with respect to
$\Omega$, which has little effect
on the physical meaning of this function but ensures that the
coupling constant
\begin{equation}
\lambda_{ph} = 2 \int_0^1 dq_x	\int_0^1 dq_y
\int_0^\infty {d\Omega\over \Omega}
\alpha^2F_{ph}({\bf q},\Omega )  = g_{ph}
\label{coupl}
\end{equation}
is independent of both $\Omega_0$ and $\Gamma$.
\cite{Manske}
The normalized ${\bf q}$-dependent form factor
 \begin{equation}
G({\bf q}) = 36.9\,e^{-25q_x^2}\,{\left( 1-q_y\right) q_y\over
1+\left( 2q_y\right)^{11}} \quad q_y>q_x
\label{form}
\end{equation}
reflects nesting properties of the Fermi surface.
%which can strongly affect the electron-phonon interaction.
The expression Eq.\ (\ref{form})
represents a good fit to the numerical results	shown in Fig. 3
of Ref.\ \onlinecite{Savrasov}.
In all our calculations the coupling constant $g_{SF}^2$ was adjusted such
that $T_c = 92\,$K was kept constant.

Details of the calculation of the self-energies  $Z({\bf k},\omega )$,
which renormalizes the frequency to $\tilde\omega  = \omega Z({\bf k},\omega )$,
$\chi ({\bf k},\omega )$, which is added to the band energy Eq.\
(\ref{qpdisp}), and of the anomalous self-energy
$\Phi ({\bf k},\omega )$
%$\Phi ({\bf k},\omega ) = \Delta ({\bf k},\omega ) Z({\bf k},\omega )$,
\cite{Bille} will have to be given elsewhere. To calculate these quantities
we used the Eliashberg equations
in the form given by Marsiglio {\it et al.} [Ref.\ \onlinecite{Marsiglio}].
%including also potential scattering in the $t$-matrix approximation.

For the energy dispersion Eq.\ (\ref{qpdisp}) with $B=0.45$
we show in Fig. 2  Re$\,Z({\bf k},\omega )$,
which is related to the mass  renormalization,
as function of $\omega$ for three
different temperatures and two points on the Fermi line, one on the zone
boundary, the other one at the position of the order parameter node.
In Fig. 3 the corresponding  results are shown for
Im$\,\omega Z({\bf k},\omega )$, which is often identified with the
quasiparticle scattering rate \cite{Hensen97}.
%Since the other self-energy parts
%are also complex the scattering rate should be determined from the pole of the
%single-particle Green function. This, however, requires calculation of the
%self-energies for arbitrary complex frequencies which is not feasible
%numerically. Approximate analytic calculations of the pole positions
%give correct results in this strong coupling theory only if the imaginary parts
%of the self-energies are very small which is usually not the case.
%Using the width of the quasiparticle spectral function to obtain
%the scattering rate is possible in the normal state. Below $T_c$, however,
%the quasiparticle peak splits into a particle and a hole contribution
%so that the interpretation of the width becomes ambiguous.

These figures show that there is considerable anisotropy already in the normal
state, indicating much stronger interactions where the Fermi line meets  the
zone boundary than on the zone diagonal.  This is due to the fact that for the
particular band considered the nesting condition ${\bf q} = (\pm 1,\pm 1)$
which maximizes the interaction Eq.\ (\ref{elisf}), is fulfilled to a much
higher degree for points ${\bf k}_F = (1.0,0.12)$ than for
${\bf k}_F = (0.37,0.37)$. Large values of $Z$ are found on the image of the
Fermi line under a translation by ${\bf Q}$, i.e. near
${\bf k} = (1-0.37,1-0.37)$.\cite{Dahm} The anisotropy of $Z$ on the Fermi
line
is not altered when phonons are included because these scatter most strongly
between Fermi lines near ${\bf k}_F = (1.0,\pm 0.12)$.
An interaction peaked at zero momentum transfer, however, would reduce the
ansitropy.

Both mass renormalization and scattering rates are smaller at low frequencies
when phonons are included because the coupling parameter $\lambda_{SF}$,
defined in analogy to $\lambda_{ph}$ Eq.\ (\ref{coupl}), has to be taken as
$\lambda_{SF} = 2.545\,$eV to obtain $T_c = 92\,$K which is larger than
$\lambda_{SF} + \lambda_{ph} = 1.979\,{\rm eV} + 0.25\,{\rm eV}$
which gives the same $T_c$. The source of this discrepancy is the contribution
to the phonon spectrum at high frequencies which is absent in $\alpha^2F_{SF}$.

When the temperature is lowered below $T_c$, the scattering rate drops much
faster at the zone boundary where it was largest. The effect that this behaviour
will have on the conductivity depends very much on the variation of the Fermi
velocity (see Fig. 1). At $T\approx 0.1T_c$, the scattering rate for $\omega <
15\,$meV has dropped well below typical disorder induced scattering rates.
\cite{Hensen97}.
\epsfxsize=3.35in
\begin{figure}[htb]
  \vspace{-1.0cm}
% \hspace{0.1cm}
  \label{fig2}
%  \begin{center}
    \centerline{%
    \begin{minipage}{8.5cm}
       \mbox{\epsffile{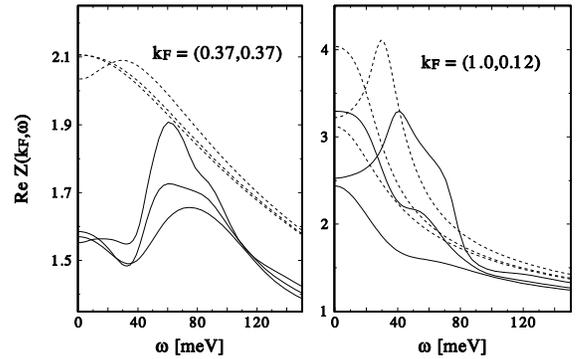}}\\
  \vspace{-0.8cm}
       \parbox{8cm}{\caption{Mass renormalization function versus frequency
for reduced temperatures $T/T_c = 1.0,\,0.9,\,0.1$. At $\omega\!=\!60\,$meV,
Re$\,Z$ increases monotonically with   decreasing temperature.
Dashed curves are for $g_{ph} = 0$ in Eq.\ ({\protect\ref{eliph}}),
solid curves are for   $g_{ph} = 0.25\,$eV.
Note the different ordinate scales.
}}
     \end{minipage}
    }
%  \end{center}
\end{figure}
\epsfxsize=3.35in
\begin{figure}[htb]
  \vspace{-1.5cm}
% \hspace{0.1cm}
  \label{fig3}
%  \begin{center}
    \centerline{%
    \begin{minipage}{8.5cm}
       \mbox{\epsffile{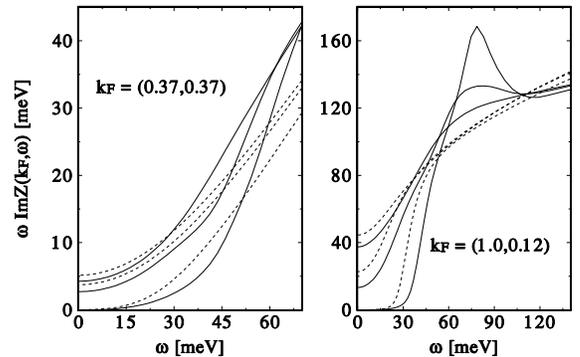}}\\
  \vspace{-0.5cm}
       \parbox{6cm}{\caption{Scattering rates versus frequency for
$T/T_c = 1.0, 0.9, 0.1$. At around $\omega = 30\,$meV, $\omega\,{\rm Im}\, Z$
increases monotonically with increasing temperature.
Dashed curves: $g_{ph} = 0$, solid curves: $g_{ph} = 0.25\,$eV. }}
     \end{minipage}
    }
%  \end{center}
\end{figure}
  \vspace{-0.3cm}
In Fig. 4 we show the order parameter
$\Delta ({\bf k},\omega ) = \Phi ({\bf k},\omega ) / Z({\bf k},\omega )$,
on the Fermi line at the zone boundary.
The ratio ${\rm Re}\,\Delta (\omega\!=\!0, T\!=\!0.1T_c)\,/\,T_c = 2.79$
without phonons and $3.42$ with phonons is larger than the weak coupling
value $2.14$ but is still at the lower end of the range of values required to
fit experiments within weak-coupling theories. \cite{Hensen97}

\epsfxsize=3.35in
\begin{figure}[htb]
  \vspace{-1.0cm}
% \hspace{0.1cm}
  \label{fig4}
%  \begin{center}
    \centerline{%
    \begin{minipage}{8.5cm}
       \mbox{\epsffile{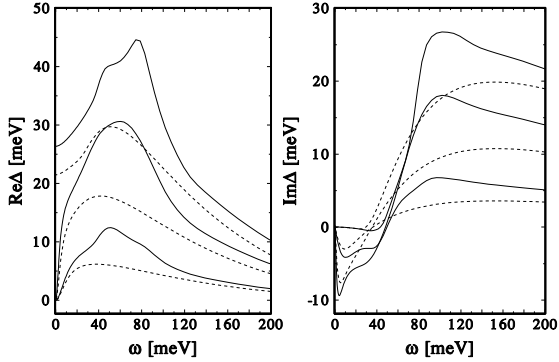}}\\
  \vspace{-0.5cm}
       \parbox{8.5cm}{\caption{Real and imaginary parts of the order parameter
versus frequency for $T/T_c = 0.99, 0.9, 0.1$  and $g_{ph} = 0$ (dashed)
and $g_{ph} = 0.25\,$eV (solid).}}
     \end{minipage}
    }
%  \end{center}
\end{figure}

The surface impedance is calculated in the local limit which is well justified
for all high-T$_c$ materials.
The conductivity is obtained without taking vertex corrections into account.
Since the interactions are momentum dependent, this is an approximation.
\cite{Pinesb}
%Spin fluctuations involve large angle scattering so that vertex corrections,
%which have the effect of projecting out forward scattering, actually reduce the
%real part of the conductivity. The phonons considered here represent small
%angle scattering so that vertex corrections resulting from both these
%scattering mechanism would partially cancel.

\epsfxsize=3.20in
\begin{figure}[htb]
  \vspace{-1.0cm}
% \hspace{0.1cm}
  \label{fig5}
%  \begin{center}
    \centerline{%
    \begin{minipage}{8.5cm}
       \mbox{\epsffile{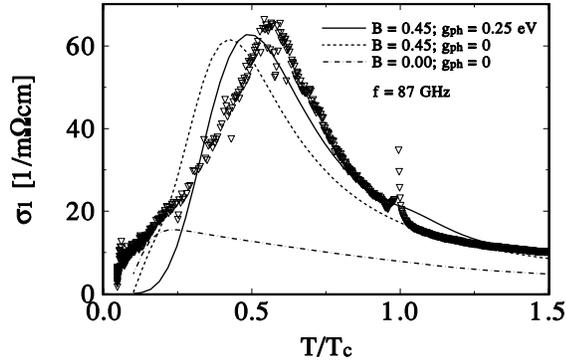}}\\
  \vspace{-0.3cm}
       \parbox{6cm}{\caption{Real part of the conductivity versus reduced
temperature at fixed microwave frequency. }}
     \end{minipage}
    }
%  \end{center}
\end{figure}
Fig. 5 shows the real part $\sigma_1$ of the conductivity at $87\,$GHz as
function of temperature for different energy dispersions and different
interactions. The only parameter adjusted in this figure is $g_{SF}$ and this is
chosen to fit $T_c$, not $\sigma_1$.
For $B=0.45$ in Eq.\ (\ref{qpdisp}) the agreement with a typical experiment
(Ref.\ [\onlinecite{Hensen97}], Sample B) is good, because the scattering rates
near the nodes are much smaller than the average value. At low temperatures,
$\sigma_1$ drops too fast because the inelastic scattering rate falls below the
experimental frequency. For these temperatures it is essential to take disorder
induced elastic scattering into account. This would also reduce the peak height
so that the conclusion seems to be inescapable that the interactions must become
smaller below $T_c$. How large a reduction one would infer from experiment
depends crucially on the energy dispersion. With $B=0$ we have nearly perfect
nesting and hence a large scattering rate on the Fermi line where the order
parameter vanishes and where the Fermi velocity is maximal. For this case we
find that the peak in $\sigma_1$ is far too small to be anywhere near the data.
\epsfxsize=3.20in
\begin{figure}[htb]
  \vspace{-1.5cm}
% \hspace{0.1cm}
  \label{fig6}
%  \begin{center}
    \centerline{%
    \begin{minipage}{8.5cm}
       \mbox{\epsffile{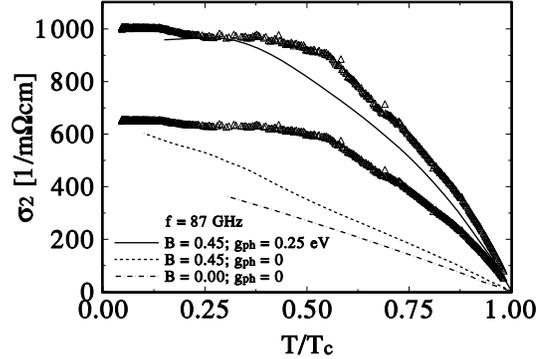}}\\
  \vspace{-0.5cm}
       \parbox{6cm}{\caption{Imaginary part of the conductivity
}}
     \end{minipage}
    }
%  \end{center}
\end{figure}

  \vspace{-0.2cm}
Fig. 6	compares the superfluid density in the form
$\sigma_2(\omega ,T)$ with data taken at $87\,$GHz.\cite{Hensen97}
Since Re$\,Z$ renormalizes the bare London penetration depth,
which is determined by the band structure, we obtain $\lambda (0) = 123\,$nm
when phonon-exchange is included instead of  $\lambda (0) = 155\,$nm.
The data are scaled to fit these different values of $\lambda (0)$.
Including phonons gives a much improved fit because of the fast rise
in Re$\,\Delta ({\bf k}_F,\omega )$
at high frequencies (see Fig. 4).
As in the case of $\sigma_1$, reaching agreement with experiment seems to
require scattering rates which drop more rapidly in the superconducting state
than our temperature-independent Eliashberg functions predict.

%\epsfxsize=3.35in
%\begin{figure}[htb]
%  \vspace{-0.5cm}
%% \hspace{0.1cm}
%  \label{fig4}
%%  \begin{center}
%    \centerline{%
%    \begin{minipage}{8.5cm}
%	\mbox{\epsffile{OA.eps}}\\
%  \vspace{-0.5cm}
%	\parbox{6cm}{\caption{Don't know yet
%}}
%     \end{minipage}
%    }
%%  \end{center}
%\end{figure}

%\epsfxsize=3.50in
%\begin{figure}[htb]
%  \vspace{-0.5cm}
%% \hspace{0.1cm}
%  \label{fig7}
%%  \begin{center}
%    \centerline{%
%    \begin{minipage}{8.5cm}
%	\mbox{\epsffile{IMZL.eps}}\\
%  \vspace{-0.5cm}
%	\parbox{6cm}{\caption{Don't know yet
%}}
%     \end{minipage}
%    }
%%  \end{center}
%\end{figure}

%\epsfxsize=3.50in
%\begin{figure}[htb]
%  \vspace{-0.5cm}
%% \hspace{0.1cm}
%  \label{fig8}
%%  \begin{center}
%    \centerline{%
%    \begin{minipage}{8.5cm}
%	\mbox{\epsffile{lambda.eps}}\\
%  \vspace{-0.5cm}
%	\parbox{6cm}{\caption{Don't know yet
%}}
%     \end{minipage}
%    }
%%  \end{center}
%\end{figure}

%\epsfxsize=3.50in
%\begin{figure}[htb]
%  \vspace{-0.5cm}
%% \hspace{0.1cm}
%  \label{fig3}
%%  \begin{center}
%    \centerline{%
%    \begin{minipage}{8.5cm}
%	\mbox{\epsffile{surimp.eps}}\\
%  \vspace{-0.5cm}
%	\parbox{6cm}{\caption{Don't know yet
%}}
%     \end{minipage}
%    }
%%  \end{center}
%\end{figure}

  \vspace{-0.5cm}
%\bibliographystyle{prsty}
%\bibliography{bostonref.bib}

\end{multicols}
\end{document}